\documentclass[11pt]{iopart}
\usepackage{iopams}
\usepackage{txfonts}
\usepackage{graphicx}
\newcommand{\fma}[1]{\mbox{$#1$}}
\newcommand{\ltsim}{\raisebox{-0.5ex}{$\;\stackrel{<}{\scriptstyle \sim}\;$}}
\newcommand{\gtsim}{\raisebox{-0.5ex}{$\;\stackrel{>}{\scriptstyle \sim}\;$}}
\newcommand{\unit}[1]{\ifmmode \:\mbox{\rm #1}\else \mbox{#1}\fi}
\newcommand{\mone}{\fma{^{-1}}}

\newcommand{\eg}{{e.g.\/}}

\newcommand{\ha}{H$\alpha$}

\newcommand{\hd}{H$\delta$}
\newcommand{\hi}{H~{\sc i}}

\newcommand{\oii}{[O~{\sc ii}]}

\newcommand{\siii}{[S~{\sc iii}]}

\newcommand{\nai}{Na~{\sc i}}

\newcommand{\caii}{Ca~{\sc ii}}

\newcommand{\kms}{\unit{km~s\mone}}

\newcommand{\msunyr}{\unit{M$_\odot$~yr\mone}}

\newcommand{\haro}{Haro\,11}
\newcommand{\esoig}{ESO\,338-IG04}
\newcommand{\esog}{ESO\,400-G43}

\begin{document}
\article[Calcium triplet kinematics in BCGs]{Star-forming Dwarf Galaxies: Ariadne's Thread in the
  Cosmic Labyrinth}{Stellar kinematics in blue compact galaxies}

\author{Robert J. Cumming$^1$, G{\"o}ran {\"O}stlin$^1$, Thomas Marquart$^2$, 
          Kambiz Fathi$^1$,
	  Nils Bergvall$^2$ and Angela Adamo$^1$ }

\address{$^1$ Department of Astronomy, Stockholm University, SE-106~91 Stockholm, Sweden.}
\address{$^2$ Department of Physics and Astronomy, Uppsala University,
     Box 515, SE-751~20 Uppsala, Sweden}
\ead{\mailto{robert@astro.su.se}
}

\begin{abstract}
In a programme of observations of local luminous blue compact galaxies
(BCGs) we are investigating kinematics by using tracers of both stars
and ionised gas. Here we summarise our program and present new data on
the local Lyman break galaxy analogue \haro. From spatially-resolved
spectroscopy around the near-infrared \caii\ triplet, we find that its
stars and ionised gas have similar velocity fields. Our programme so
far indicates however that emission line velocities can differ locally
by a few tens of \kms\ from the \caii\ values. Comparing our data to
simple stellar population models, we assess which stellar population
the \caii\ triplet traces and its potential beyond the local universe.
\end{abstract}



\section{Introduction \label{intro}}

Two central questions in galaxy evolution are how
starbursts are triggered and how they are quenched, and how
triggering and quenching are related to galaxy properties and
environments. 
Starburst episodes in easily observable, nearby 
galaxies may be started by collisions with other galaxies and/or gas
clouds, but a galaxy's star formation rate is also likely to vary
naturally during its history, simply as a result of internal motions
and feedback. Kinematic studies which include both gas and stars have
the potential to distinguish between triggering mechanisms.

Blue compact galaxies (BCGs) are a promising population for
investigating these effects. They tend to be only moderately affected
by internal reddening, allowing us a fairly unencumbered view of the
star-forming region. BCGs in the local universe ($D \ltsim$100 Mpc)
span the range between star-forming dwarfs and systems whose
properties more closely resemble starbursting galaxies at higher
redshift (\eg\ Lyman break galaxies \cite{Giavalisco}; LBG) while
still being close enough for spatially resolved spectroscopy.  \haro\
at $D$=88 Mpc is one of the best local LBG analogues
\cite{Grimes}. Its high star formation rate ($\sim$20 \msunyr
\cite{o01}) and lack of detected \hi\ \cite{bo02} suggests the starburst
may be running out of gas.

{\"O}stlin \etal \cite{o01} observed and analysed the ionised gas
kinematics of a sample of BCGs using Fabry-Perot spectroscopy in
\ha. The picture that emerged of galaxies with predominantly disturbed
kinematics and absence of rotational support suggested that
large-scale dynamical disturbances --- mergers, for example --- could
have triggered their current starbursts. Nevertheless, feedback from
supernova explosions and stellar winds may dominate the emission-line
kinematics and compromise our ability to distinguish large-scale
dynamical effects.

We are carrying out a programme of observations of blue compact
galaxies with the aim of measuring and comparing kinematics in both
stars and ionised gas. To trace the stars we use the near-infrared
\caii\ triplet (CaT), and for the gas, \siii\ and Paschen emission
lines. Our lines are close in wavelength and avoid differential
interstellar extinction.
Other potential stellar tracers are either contaminated by
interstellar absorption (\nai\,D, \caii\ H, K; \cite{mck}) or too weak
(Mg~{\sc i} $\lambda$5170).

In our programme so far we have reported results for three \esog\
\cite{o04}, He\,2-10 \cite{m07} and \esoig\ \cite{c08}. Here we
summarise the program so far, present early results for \haro\
(ESO\,350-IG38) and take a critical look at the CaT method and which
stars it traces.

   \begin{figure}
   \centering
   \includegraphics[width=8.3cm]{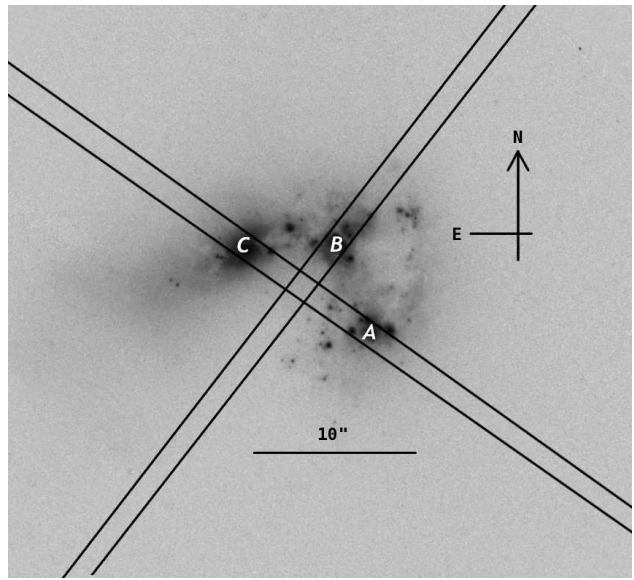}

      \caption{HST F550M image of \haro\ \cite{o07} showing our two 0.7-arcsec slits and the
      three main emission knots \cite{Vader93}. 10 arcsec
      corresponds to 3.8 kpc. }

         \label{f-slits}
   \end{figure}

   \begin{figure}
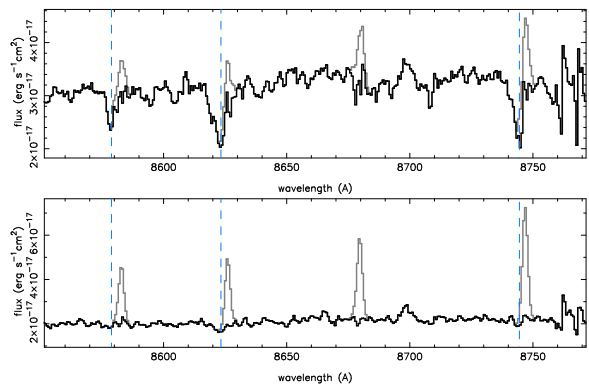

   \centering
   \includegraphics[angle=270,width=.49\textwidth]{Cumming_Fig2a.ps}\\
   \includegraphics[angle=270,width=.49\textwidth]{Cumming_Fig2b.ps}

      \caption{Examples of Paschen-line subtraction from the
      spectrum of \esoig\ \cite{c08} around the calcium triplet, and
      how its success depends on the relative strengths of the
      Paschen-line emission and the \caii\ absorption. The subtracted
      spectrum is plotted over the original spectrum (grey).  Vertical
      dashed lines mark the approximate expected position of the
      centres of the three calcium triplet lines. }

         \label{pasub}
   \end{figure}

   \begin{figure*}
   \centering
   \includegraphics[angle=-90,width=.88\textwidth]{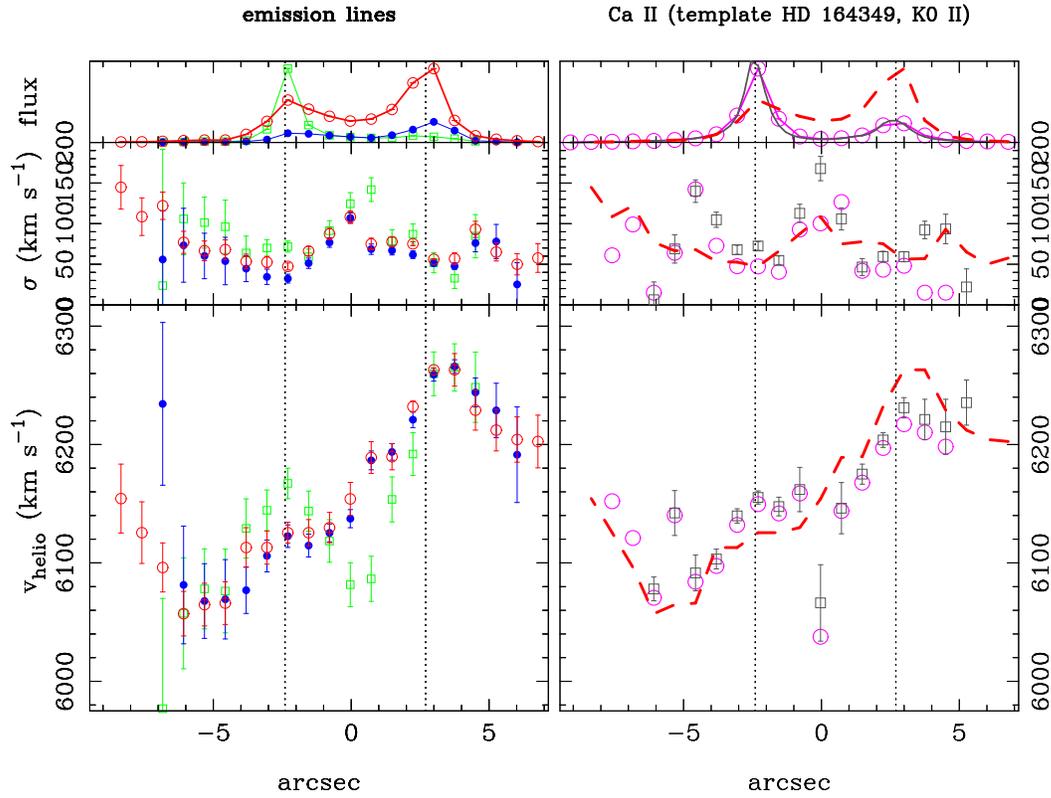}

   \caption{For \haro, variation of continuum/line flux, velocity
   dispersion $\sigma$ and heliocentric velocity, for the gas (left)
   and stars (right) along slit PA 237.5$^\circ$-57.5$^\circ$.  Dotted
   vertical lines mark the positions of knots C (left) and A (right;
   see also Fig.\ \ref{f-slits}). Zero on the abscissa is at the
   \ha\ kinematical centre of the galaxy \cite{o01}. Spectral resolution was 63 \kms\ and seeing around 0.7 arcsec.

   The left panel shows the emission lines \siii\ $\lambda$9069 (red
   open circles), \hi\ Paschen lines (weighted means of all measured
   lines; blue filled circles), and O~{\sc i} $\lambda$8446 (green
   squares).  The flux scales show (left) the normalised strengths of
   the emission lines and (right) the \siii\ line flux compared to the
   mean continuum value around the calcium triplet.   Knot C shows remarkably strong O~{\sc i},
   perhaps due to dense gas retained in its central SSC after shell
   ejection (\eg\ J.\ Palous, these proceedings).

   The right panel shows results of cross-correlation with the K0
   bright giant HD\,164349 (squares) and pPXF results (open
   circles). The formal errors in the measured $v_{\rm helio}$ and
   $\sigma$ given by the cross-correlation software (fxcor in IRAF)
   underestimate the systematic errors (see text).  Cross-correlation
   measurements of velocity dispersion and heliocentric velocity are
   only shown where the CCF peak is greater than 0.25. CCF and pPXF
   results are not shown where the Paschen-subtraction quality was
   deemed to be poor \cite{c08}. The thick dashed line in the right
   panel repeats the \siii\ results from the left panel for
   comparison.  10 arcsec corresponds to 3.8 kpc at the distance of
   \haro.}

      \label{rc}
   \end{figure*}
   \begin{figure}
   \centering
   \includegraphics[angle=-90,,width=.9\textwidth]{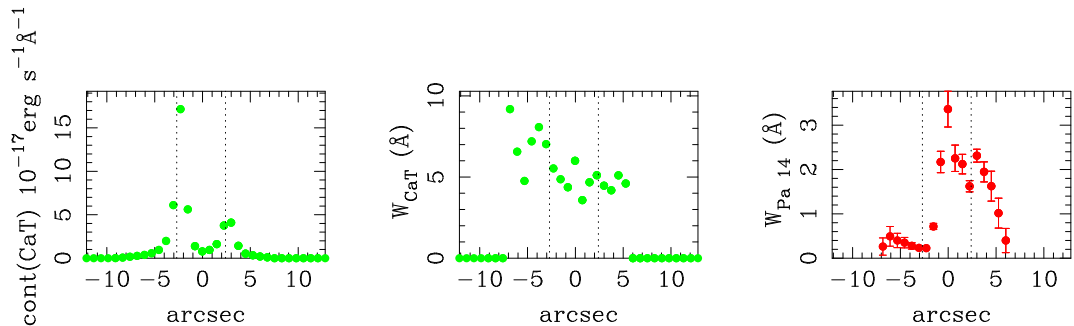}

      \caption{Spatial variation across \haro\ of (left) the continuum
      around the calcium triplet, (centre) the combined CaT equivalent
      width (defined according to ref.\ \cite{Cenarro01}), and (right) the
      equivalent width of Pa\,14 $\lambda$8598 which needs to be low
      if we are to reliably trace stellar velocities using CaT.}

         \label{contrast}
   \end{figure}

\section{Observations and method}
Our observations are presented in detail in our previous papers
\cite{o04,m07,c08}. Here we introduce long-slit spectra of \haro\
taken using FORS2 at the VLT on 2003 Aug 6-7, to be presented in
Cumming \etal\ (in preparation). 
The galaxy spectra were complemented by observations of six bright
late-type giants and supergiants which we use as templates for
cross-correlation.

After extracting spatially-separated spectra, we modelled and subtracted
the Paschen line emission which contaminates the red wing of all three
\caii\ triplet lines (Fig.\ \ref{pasub}). This procedure is critical
if we are to derive correct velocities. If stars and gas follow each
other, for example, oversubtraction will give a spuriously high
velocity, and undersubtraction a CaT line that is too blue. The Paschen
line model was constructed using as templates the strongest clean
Paschen lines (Pa\,10+11+12), or where they are too weak, the strong
\siii\ $\lambda9069$ line. Success requires good signal to noise ratio
both in the observed Paschen lines and in the stellar continuum.

To recover the stellar velocity and an estimate of the velocity
dispersion from the subtracted spectra, we use two different methods:
cross-correlation with template stars \cite{HF96a,HF96b,NW} and penalised
pixel-fitting (pPXF) using a template library \cite{CE}. We have
described our implementation of these procedures in detail in ref.\
\cite{c08}.

\section{Results} \label{results}

\subsection{Star and gas velocity fields}
We present our spatially-resolved stellar and gas velocity
measurements for \haro\ in Figure \ref{rc}. Our slit joins knots A and
C (Fig.\ \ref{f-slits}).  The velocities measured in the gas and the
stars follow one another to within the likely experimental errors. The
stellar velocity dispersions are likewise similar to the velocity
dispersions measured from the emission line widths.

The stellar velocity measurements are however strongly affected by
imperfect Paschen-line subtraction, particularly where the contrast
between the stellar lines and Paschen emission is large.  Where the
Paschen lines are strong and the stellar signal weak, the stellar
velocities are more uncertain than the error bars indicate.

\subsection{Line equivalent widths}
Our spectra provide some other clues to the stars whose velocities we
are tracing. In Figure \ref{contrast} we plot three measures of the
stellar population in \haro: (1) the strength of the continuum close
to the \caii\ triplet, which we expect to be dominated by stars
\cite{tdt}; (2) the equivalent width of CaT, $W_{\rm CaT}$; and (3)
$W_{\rm Pa 14}$ (see sect.\ \ref{pop}), the equivalent width of the
unblended Paschen line between \caii\ $\lambda\lambda8542$ and 8662
(scaling from the flux in Pa\,10 to ensure suitable signal-to-noise
ratio).

The scatter in $W_{\rm CaT}$ is large but is noticeably larger around
knot C ($\sim$7~{\AA}) than knot A ($\sim$5~{\AA}). $W_{\rm Pa 14}$ is very
low around knot C and peaks around 2~{\AA} in the nebulosity surrounding
knot A. Our other slit shows an even stronger peak in $W_{\rm Pa 14}$
at knot B.

\section{Discussion}

\subsection{Stars and gas follow each other, but not always}
Our programme so far indicates that CaT absorption in BCGs in some
cases follows the emission-line kinematics (\esoig, \haro; \cite{c08})
but in others shows differences of up to a few tens of \kms\ (\esog,
He\,2-10; \cite{o04,m07}). This indicates that bright emission lines
are unreliable tracers of the stellar velocities in BCGs. What this
tells us about the dynamics of each galaxy and feedback processes
depends, however, on the stellar population whose kinematics we are
measuring.

\subsection{Spatial resolution and stellar populations}\label{pop}
We have compared our equivalent width measurements (Fig.\ \ref{contrast}) with stellar
population models in an attempt to understand which stars the \caii\
triplet allows us to trace, and which stars the Paschen emission lines
conceal from our view.

   \begin{figure}
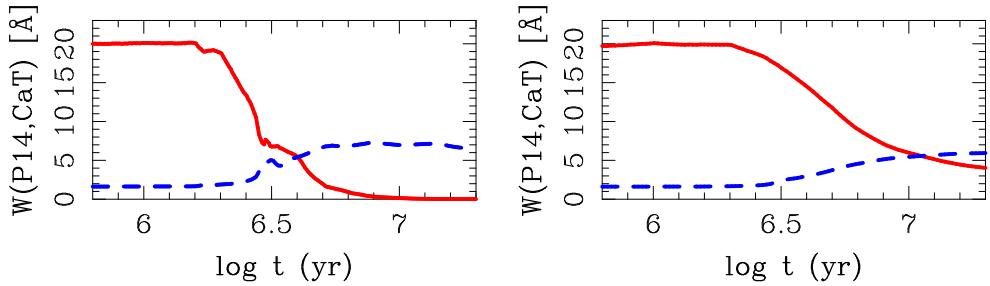

   \centering
   \includegraphics[angle=-90,width=.4\textwidth]{Cumming_Fig5a.ps}\hspace{0.5cm}\includegraphics[angle=-90,width=.4\textwidth]{Cumming_Fig5b.ps}

      \caption{Evolution of the equivalent widths of CaT absorption
      (dashed line) and Pa\,14 in emission (solid line) according to
      Starburst99 model runs designed to match \haro, with $Z = 0.004$
      and an instantaneous burst (left) or continuous star formation
      (right). The models assume that the ionising population and the
      ionised gas are coextensive. For instantaneous bursts, CaT and
      the Paschen lines sample populations with different ages.}

      \label{ew}
   \end{figure}

Reliable CaT velocities require low equivalent widths in the Paschen
series. Our long-slit data require $W_{\rm Pa 14}\ltsim
2$~{\AA}. This puts constraints both on the stellar population and, more
practically, how well we can resolve the galaxy, since we need
unencumbered line of sight to the population in question. We have
used the Starburst99 population synthesis code
\cite{Leitherer} to see how the equivalent widths of Pa\,14 and CaT
can be expected to evolve in a starburst galaxy like \haro\ (Fig.\
\ref{ew}).

In a model with constant star formation rate (Fig.\ \ref{ew}, right
panel), it takes well over $10$ Myr for the Paschen lines to fade
sufficiently. This situation is appropriate if our resolution element
include local instantaneous bursts at different evolutionary
stages. It seems likely that \haro's star formation was spread out
over time \cite{bo02}, so integrating the light of the whole galaxy will
probably make it look like a continuous starburst.
However, our spatial resolution allows us to resolve smaller regions
(\eg\ knots A, B and C) which may be modelled as instantaneous bursts
(Fig.\ \ref{ew}, left panel). After only 5 Myr, recombination
emission from such a burst has dropped enough for CaT to be
measurable.  At such ages, we expect the CaT to be mostly provided by
red supergiants (RSGs) \cite{mayya}; in older populations, CaT
absorption comes instead from low-mass red giants.  Low-metallicity
galaxies like \haro\ the contribution to $W_{\rm CaT}$ from RSGs is
expected to be lower than in high-metallicity systems, and this
accounts for the lack of an RSG peak in the evolution of $W_{\rm
CaT}$.

In \haro, knot C and the region surrounding it provides the best
measurements of CaT absorption. In knot C, while we see no evidence of
significant Paschen absorption lines, the Balmer series is present in
absorption and can be used as an age estimator.
We have used the absorption wings of \hd\ and H8 to measure their
equivalent widths, and compared with models from ref.\ \cite{gdl} (as
we did in ref.\ \cite{ocb}), finding an age of about 20 Myr.
%
In knots A and B, the lack of Balmer absorption and high emission
equivalent widths indicate instead ages of only a few Myr. The
starburst in \haro\ appears to have been spread out over at least the last
$\sim$20 Myr, and we can reliably trace the stellar velocities in CaT
only where the older population is not outshone by the younger
gas-rich regions. An intermediate case is where super star clusters
(Adamo, these proceedings) have ejected their ionised gas and can be
distinguished from the surrounding wind-blown bubble (\eg\ cluster \#23 in \esoig\ \cite{ocb}). 
%

CaT kinematic data like ours have the potential to separate the
kinematics of stars and gas in at least nearby star-forming galaxies
and distinguish between models of how they form and
evolve. Simulations have indicated that mergers can account for the
disturbed morphologies seen in BCGs \cite{Jesseit,bekki}. Bekki
\cite{bekki} in particular finds that new stars form close to the gas,
presumably with similar kinematics. Our results for new stars in real
BCGs are broadly consistent with this.

\subsection{Implications for galaxies at higher redshift}
In high-redshift galaxies, the velocity dispersion and shear in strong
emission lines may be the only kinematic observables (\eg\ Erb \etal
\cite{Erb06}). The relationship between emission-line widths and for
example dynamical masses \cite{Pettini01} is important to anchor in a
low-redshift context with galaxies similar to those we can see at
earlier epochs.  Kobulnicky \& Gebhardt \cite{KG} observed a sample of
nearby late-type galaxies (including the BCG He\,2-10) in \hi, \caii\
H and K and \oii, smearing using telescope drift to estimate each
source's total spectrum, and concluded that nebular emission lines
generally agree with the other tracers.

Our data indicate that in BCGs, emission line kinematics are probably
no better than a first approximation to the stellar velocity
field. Moreover, while the \caii\ triplet traces the stellar
population nicely, reliable results require either very high
signal-to-noise ratio or resolved regions with small equivalent widths
in the competing Paschen emission lines (ages $\gtsim$8 Myr; Fig.\
\ref{ew}). At moderate redshifts, this problem will exacerbated by
limited spatial resolution and alternative tracers of stellar velocity
should be investigated.

\ack Thanks to Matthew Hayes, Genoveva Micheva, Casiana Mu{\~n}oz-Tu{\~n}on, Polis Papaderos, Eduardo Telles, Guillermo Tenorio-Tagle, Pepe Vilchez and Erik
Zackrisson for discussions.


\section*{References}
\begin{thebibliography}{120}
\bibitem{Giavalisco} Giavalisco, M 2002, ARAA {\bf 40} 579
\bibitem{Grimes} Grimes, J~P \etal\ 2007, ApJ {\bf 668} 891
\bibitem{bo02} Bergvall, N, {\"O}stlin, G 2002, A\&A {\bf 390} 891
\bibitem{o01} {\"O}stlin, G, Amram, P, Bergvall, N, Masegosa, J, Boulesteix, J, M{\'a}rquez, I 2001, A\&A {\bf 374} 80
\bibitem{mck} McKeith, D, Castles, J, Greve, A, Downes, D 1993, A\&A {\bf 272} 98
\bibitem{o04} {\"O}stlin, G, Cumming, R~J, \etal\
2004, A\&A {\bf 419} L43
\bibitem{m07} Marquart, T, Fathi, K, {\"O}stlin, G, Bergvall, N, Cumming, R~J, Amram, P 2007, A\&A {\bf 474} L9
\bibitem{c08} Cumming, R~J, Fathi, K, Marquart, T, {\"O}stlin, G, Bergvall, N, Masegosa, J, M{\'a}rquez, I, Amram, P 2008, A\&A {\bf 479} 725
\bibitem{o07} {\"O}stlin, G \etal\
2008, AJ, in press (arXiv:0803.1174)
\bibitem{Vader93} Vader, J~P, Frogel, J, Terndrup, D~M, Heisler, C~A 1993, AJ {\bf 106} 1743
\bibitem{Cenarro01} Cenarro, A~J, Cardiel, N, Gorgas, J, Peletier, R~F, Vazdekis, A, Prada, F 2001, MNRAS {\bf 326} 959
\bibitem{HF96a} Ho, L, Filippenko, A~V 1996, ApJ {\bf 466} 83
\bibitem{HF96b} Ho, L, Filippenko, A~V 1996, ApJ {\bf 472} 600
\bibitem{NW} Nelson, C~H, Whittle, M 1995, ApJS {\bf 99} 67
\bibitem{CE} Cappellari, M, Emsellem, E 2004, PASP {\bf 116} 138
\bibitem{Leitherer} Leitherer, C \etal\ 1999, ApJS {\bf 123} 3
\bibitem{tdt} Terlevich, E, D{\'\i}az, A, Terlevich, R 1990, MNRAS {\bf 242} 271
\bibitem{mayya} Mayya, Y~D 1997, ApJ {\bf 482} 149
\bibitem{dtt} D{\'\i}az, A~I, Terlevich, E, Terlevich, R 1989, MNRAS {\bf 239} 325
\bibitem{Vega} Vega, L~R, Asari, N~V, Cid Fernandes, R, Garcia-Rissmann, A, Storchi-Bergmann, T, Gonz{\'a}lez Delgado, R~M, Schmitt, H 2008, MNRAS, in press (arXiv:0809.3178)
\bibitem{gdl} Gonz{\'a}lez Delgado, R~M, Leitherer, C 1999, ApJS {\bf 125} 489
\bibitem{ocb} {\"O}stlin, G, Cumming, R~J, Bergvall, N, 2007, A\&A {\bf 461} 471
\bibitem{Jesseit} Jesseit, R, Naab, T, Peletier, R~F, Burkert, A 2007, MNRAS {\bf 376} 997
\bibitem{bekki} Bekki, K 2008, ApJ {\bf 680} 29
\bibitem{Erb06} Erb, D, Steidel, C, Shapley, A~E, Pettini, M, Reddy, N~A, Adelberger, K 2006, ApJ {\bf 646} 107
\bibitem{Pettini01} Pettini, M, \etal\
2001, ApJ {\bf 554} 981
\bibitem{KG} Kobulnicky, H~A, Gebhardt, K 2000, AJ {\bf 119} 1608


\end{thebibliography}
\end{document}